\documentclass[]{aa}
\usepackage{epsfig}
\usepackage{rotating}
\newcommand{\sub}[1]{_{\rm #1}}
\newcommand{\reference}[1]{}
\newcommand{\oversim}[2]{\protect{\mbox{\lower0.5ex\vbox{%
   \baselineskip=0pt\lineskip=0.2ex
   \ialign{$\mathsurround=0pt #1\hfil##\hfil$\crcr#2\crcr\sim\crcr}}}}} 
\newcommand{\sig}{\:\lower0.6ex\hbox{$\stackrel{\textstyle >}{\sim}$}\:}
\newcommand{\sil}{\:\lower0.6ex\hbox{$\stackrel{\textstyle <}{\sim}$}\:}
\newcommand{\sigs}{\:\lower0.4ex\hbox{$\stackrel{\scriptstyle
      >}{\scriptstyle \sim}$}\,}
\newcommand{\sils}{\:\lower0.4ex\hbox{$\stackrel{\scriptstyle
      <}{\scriptstyle \sim}$}\,}
\renewcommand{\S}{Sect.\ }
\newcommand{\changed}{}
\def\apj{{\em ApJ }}
\def\apjs{{\em ApJS }}
\def\aap{{\em A\&A }}
\def\mnras{{\em MNRAS }}
\def\prl{{\em Phys.\ Rev.\ Lett.\ }}
\def\jcp{{\em J.\ Comp.\ Phys.\ }}

\def\pasj{{\em PASJ }}

\def\araa{{\em ARA\&A }}
\begin{document}
\title{On the structure of self-gravitating molecular clouds}
\subtitle{}
\titlerunning{The structure of self-gravitating clouds}

\author{V.\ Ossenkopf\inst{1}, 
        R.~S.\ Klessen\inst{2,3} \and \ F.\ Heitsch\inst{3}}
\authorrunning{V.\ Ossenkopf, R.~S.\ Klessen, F.\ Heitsch}

\institute{
1. Physikalisches Institut, Universit\"at zu K\"oln,
Z\"ulpicher Stra\ss{}e 77, 50937 K\"oln, Germany (e-mail: ossk@ph1.uni-koeln.de)
\and
UCO/Lick Observatory, University of California, Santa Cruz, 
CA 95064, USA (e-mail: ralf@ucolick.org)
\and
Max-Planck-Institut f{\"u}r Astronomie,
K{\"o}nigstuhl 17, 69117 Heidelberg, Germany (e-mail: heitsch@mpia-hd.mpg.de)
}
\date{Received: July 16, 2001/ Accepted: September 17, 2001}

\abstract{
To study the interaction of star-formation and turbulent molecular cloud structuring,
we analyse numerical models and observations of self-gravitating
clouds using the $\Delta$-variance as statistical measure for structural
characteristics. In the models we resolve the transition
from purely hydrodynamic turbulence to gravitational collapse
associated with the formation and mass growth of protostellar
cores. We compare models of driven and freely decaying
turbulence with and without magnetic fields.
Self-gravitating supersonic turbulence always produces a density structure
that contains most power on the smallest scales provided by 
collapsed cores as soon as local collapse sets in. This 
is in contrast to non-self-gravitating hydrodynamic turbulence where 
the $\Delta$-variance is dominated by large scale structures. \\
To detect this effect in star-forming regions observations have
to resolve the high density contrast of
protostellar cores with respect to their ambient molecular cloud. 
Using the 3mm continuum map of a star-forming cluster in Serpens we show that
the dust emission traces the full density evolution.
On the contrary, the density range accessible by molecular line observations
is insufficient for this analysis. Only dust emission and dust
extinction observations are able to to determine the structural 
parameters of star-forming clouds following the density 
evolution during the gravitational collapse.
}

\maketitle

\section{Introduction}
\label{sec:intro}

Stars form through gravitational collapse in clouds of
molecular hydrogen. Within these clouds random supersonic gas motions
associated with interstellar turbulence lead to transient
density fluctuations, some of which may exceed the
threshold of gravitational stability and collapse to form stars (e.g.\ 
\cite{elm93}, \cite{pad95}, \cite{pad99}). 

Detailed studies of the hydrodynamic and magnetohydrodynamic collapse
in molecular clouds by \cite{KHM} and \cite{HMK} show that the
location, time scale and efficiency of star formation are intimately
connected to the stochastic properties of the underlying turbulent
velocity field. To derive a consistent picture of the processes
from turbulent fragmentation to star formation it is important to 
understand how
the interplay between supersonic turbulence and self-gravity shapes
the density and velocity structure of star-forming clouds. Thus we
need a quantitative measure for the evolution of the turbulence
structure during gravitational collapse. The density and velocity
structure have to be compared on different scales to follow the
process of structure transformation from the molecular cloud size to the
scales of collapsed cores and to understand the influence of gravity
on the turbulent driving and the overall cloud structure.

A variety of different statistical measures have been deployed to
analyse the structure of turbulence either in observed molecular
clouds or in numerical simulations. Many of them are summarised in the
reviews by \cite{Semadeni2000} and \cite{Ossenkopf2000}.  One
technique especially sensitive to the discrimination of the relative
structural variation on different spatial scales is the
$\Delta$-variance introduced by Stutzki et al.\ (1998). It provides a
good separation of noise and observational artifacts from the real
cloud structure and for isotropic systems its slope is directly
related to the spectral index of the corresponding power
spectrum. \cite{Bensch} and \cite{MLO} have shown that it can be
applied in an equivalent way both to observational data and
hydrodynamic and magneto-hydrodynamic turbulence simulations allowing
a direct comparison.

Their investigations, however, neglect the influence of gravitational
collapse on the structure formation so that their conclusions may be
limited when applied to star-forming regions. It is essential to
include the effects of self-gravity for the analysis of star-forming
regions. It is the aim of the current study to close this gap
and investigate the interaction between turbulence and self-gravity.
We apply the
$\Delta$-variance to characterise the structure in numerical models of
driven and decaying self-gravitating supersonic (magneto-)hydrodynamic
turbulence and compare the results to observed regions of star
formation.

In \S\ref{sec:numerics} we
introduce the molecular cloud models analysed here and the numerical
methods used for their generation. In \S\ref{sec:results} we apply
the $\Delta$-variance analysis and discuss the time evolution 
of the structure as local collapse sets in
and protostellar cores form and accrete mass. Using various models
and comparing the density and velocity structure we demonstrate how
power is build up at the different scales during star formation. 
\S\ref{sec:observations} provides a comparison with observational 
data including dust emission and molecular lines and discusses the 
resulting differences.
We summarise our results in \S\ref{sec:summary}.

\section{Turbulence Models}
\label{sec:numerics}

\begin{table*}
\caption{\label{tab:models-2}
Properties of the considered turbulence models together with the resulting time
scales}
\begin{center}
\begin{tabular}[h]{llcccrr}
model &
description  &
$k_{\rm d}$$^{a}$ &
numerical method &
further reference$^{b}$ &
$\tau\sub{5\%}^{c}$ &
$f\sub{\tau\sub{ff}}^{d}$\\
\hline
S01 & driven HD turbulence & $1\dots2$ &
SPH & ${\cal B}1h$ in KHM  & 0.6     & 28\%\\
Sd1 & decaying HD turbulence & $1\dots2$ &
SPH & ---                  & 0.6     & 60\%\\
S02 & driven HD turbulence & $7\dots8$ &
SPH & ---                  & $>5.5$  & 0.6\%\\
Sd2 & decaying HD turbulence & $7\dots8$ &
SPH & ---                  & 2.0     & 0.0\% \\
G   & Gaussian density     & --- &
SPH & $\cal I$ in KB       & 1.3     & 9\%\\
H01 & driven HD turbulence & $1\dots2$ & 
ZEUS & ${\cal D}1h$ in KHM & 0.6     & 24\%\\
H02 & driven HD turbulence & $7\dots8$ &
ZEUS & ${\cal D}3h$ in KHM & 4.7     & 0.5\% \\
M01 & driven MHD turbulence & $1\dots2$ & 
ZEUS & ${\cal E}h1i$ in HMK& 1.2     & 6\%  \\
\hline\\
\end{tabular}
\end{center}

$^{a}${Wavenumber of the original driving}\\ $^{b}${Model names 
in the original papers: KB -- Klessen \& Burkert (2000),
KHM -- Klessen et al.\ (2000), HMK -- Heitsch et al.\ (2001)}\\
$^{c}${Time at which 5\% of the total mass is accreted onto cores
in internal units where $\tau_{\rm ff}=1.5$.}\\
$^{d}${Mass fraction that is accreted onto cores after one global
free-fall time.}\\
\end{table*}

The large observed linewidths in molecular clouds imply the presence
of supersonic velocity fields that carry enough energy to
counterbalance gravity on global scales (Williams et al.\ 2000). 
As turbulent energy dissipates rapidly, i.e.\ roughly on the free-fall
time scale (Mac Low et al.\ 1998, Stone et al.\ 1998, Padoan \&
Nordlund 1999), either interstellar turbulence is continuously
replenished in order to prevent or considerably postpone global
collapse, or alternatively, molecular clouds evolve rapidely and never
reach dynamical equilibrium between kinetic energy and self-gravity
(Ballesteros-Paredes et al.\ 1999a, Elmegreen 2000). 
{\changed Supersonic turbulence may be driven by Galactic rotation 
and supernovae on large scales and by outflows from young 
stellar objects on small scales (see e.g. \cite{Avillez}).} %RSK
The current models investigate both cases, continuously replenished
and freely decaying turbulence. 

We select a set of numerical models mostly from preexisiting studies
that spans a large range of the parameter space relevant for 
Galactic molecular clouds. We analyse the time evolution of their
density and velocity structure as gravitational
collapse progresses. Altogether, we include eight models
summarised in Tab.\ \ref{tab:models-2}.  They differ in the scale on
which turbulent driving occurs, the persistence of this driving, the
inclusion of magnetic fields, and the numerical algorithm used to
solve the hydrodynamic or magnetohydrodynamic equations.

In the present analysis we neglect feedback effects from the
produced young stellar objects (like bipolar outflows, stellar winds,
or ionising radiation from new-born O or B stars).  Thus our results
will hold only for early stages of star-forming systems.

\subsection{The driving process}

To represent hydrodynamic turbulence which carries most energy on large scales
we consider a freely decaying (model Sd1) and a driven case (S01).
Turbulence that is driven at large wavelength leads to a clustered
mode of star formation (Klessen et al.\ 2000) and appears to yield the
most appropriate description of molecular cloud dynamics (see e.g.\
Ossenkopf \& Mac Low 2001).  Furthermore, we consider decaying and
driven small-scale hydrodynamic turbulence (models Sd2 and S02,
respectively).

In all driven models the energy input is adjusted to yield the same
turbulent Mach number ${\cal M}_{\rm rms} = 10$. The Mach number is
calculated from the three-dimensional rms velocity dispersion, ${\cal
M}_{\rm rms} \equiv v_{\rm rms}/c_{\rm s} = \sqrt{2E_{\rm kin}/M}/c\sub{s}$,
where $c_{\rm s}$ is the sound speed, $E_{\rm kin}$ the total kinetic 
energy and $M$ the total mass in the system. 
Turbulent equilibrium is obtained applying a non-local
driving scheme, that inserts energy in a limited range of wavenumbers
such that the total kinetic energy contained in the system remains
constant and compensates the gravitational contraction on global
scales (Mac~Low 1999).

For the decaying models this Mach number holds at times $t \le 0$, before
turbulence is allowed to decay. The driving stops when gravity is 
``turned on'' at $t=0$ when the stage of fully established turbulence is 
reached.  We also consider a case of large-scale
contraction from Gaussian density fluctuations without any turbulent
support (model G).  This represents the most extreme case of clustered
star formation (Klessen 2001).

\subsection{The numerical implementation}

All models are computed either using SPH ({\em smoothed particle hydrodynamics}), 
a particle-based Lagrangian scheme to solve the equations of hydrodynamics,
or ZEUS, a grid-based code. We
directly compare the two codes in the case of driven turbulence. Here
we accompany the SPH models S01 and S02 by the corresponding ZEUS models
H01 and H02. This has proven to be a very effective means of
distinguishing physical results from numerical artifacts
(Mac~Low et al.\ 1998), as these two approaches practically bracket the
real dynamical behaviour of interstellar turbulence.  The density
enhancements are spatially somewhat too small and rigid in the SPH model
and too extended in the grid-based compuations (discussed in
detail by Klessen et al.\ 2000).

In SPH the fluid
is represented by an ensemble of particles, where thermodynamical
observables are obtained as local averages (Benz 1990, Monaghan
1992). The method is very flexible and can resolve high density
contrasts by increasing the particle concentration where needed.  We
use SPH in combination with the special-purpose hardware device GRAPE
(Sugimoto et al.~1990, Ebisuzaki et al.~1993), which allows
calculations at supercomputer level on a normal workstation, and adopt
periodic boundary conditions (Klessen 1997) focusing on the evolution
of a sub-region within a larger, globally stable molecular cloud.

The simulations use $2\,10^5$ SPH particles for all turbulence models
except the collapse of the Gaussian density fluctuations where we
use $5\,10^5$ SPH particles.  Shock-generated density fluctuations
with masses exceeding the local Jeans limit become unstable and
collapse. Once their central nuclei exceed a density contrast of
$10^5$, they are identified as protostellar cores and replaced by
``sink'' particles (Bate et al.\ 1995). These inherit the combined
masses, linear and ``spin'' angular momenta of the replaced particles,
and have the ability to accrete further SPH particles from their
infalling gaseous envelopes. This technique allows us to follow the
dynamical evolution of turbulent molecular cloud regions over many
global free-fall time scales. The numerical scheme and the 
scaling properties of the model are discussed in detail by Klessen \&
Burkert (2000) and Klessen et al.~(2000).

For the grid-based simulations 
we use the astrophysical MHD code ZEUS-3D (Clarke 1994). 
This is a three-dimensional version of the code described by 
Stone \& Norman (1992a,b) using second-order advection (van Leer 
1977) and von Neumann artificial viscosity introduced to resolve 
shocks. Moreover, structures close to the grid resolution 
are subject to numerical dissipation as discussed by Mac~Low \&
Ossenkopf (2000). 

The simulations were performed on a three-dimensional, uniform,
Cartesian grid containing 256$^3$ cells and assuming periodic boundary
conditions. All quantities are scaled to match the corresponding
definitions in the SPH simulations (see also Heitsch et al.\ 2001).

\subsection{Magnetic fields}

To test the influence of magnetic fields we consider the driven
turbulence model which carries most energy on large scales and
add  magnetic fields to the ZEUS simulations. By comparing
the resulting model M01 with the hydrodynamic model H01 we get
a direct measure for the significance of magnetic fields 
for the interplay of turbulence and self-gravity in  structure
formation. 

The magnetic field in this model is selected to be a {\changed major
factor where} the ratio between thermal and magnetic pressure $\beta =
p\sub{th}/p\sub{mag} = 8 \pi c\sub{s}^2\rho/B^2 = 0.04$.
With a turbulent Mach number of ${\cal M}_{\rm rms} = 10$ we
find that the turbulent velocity dispersion exceeds the Alfv\'{e}n 
speed by a factor of 1.4 so that the
structure is still essentially determined by supersonic turbulence
and only modified by the magnetic field. The mass in the cloud
still exceeds the critical mass for a  magnetostatically stable
cloud by a factor 2 (Heitsch et al. 2001) so that the field should not
prevent gravitational collapse. Cases of sub-Alfv\'{e}nic
non-self-gravitating turbulence where the whole structure is 
dominated by the magnetic field have been discussed by Ossenkopf
\& Mac~Low (2001).

\subsection{Scaling}

The models presented here are computed in normalised units. 
Throughout this paper we give all size values relative to the total 
cube size, all density values relative to the the average density
in the cube, and all velocities relative to the sound speed.
Model G contains 220 thermal Jeans masses, wherease all other models 
have 120 thermal Jeans masses\footnote{Throughout this paper the
  spherical definition of the Jeans mass is used, $M_{\rm J} \equiv
  4/3\,\pi \rho \lambda_{\rm J}^3$, with density $\rho$ and Jeans
  length $\lambda_{\rm J}\equiv \left(\pi{\cal R}T /G
      \rho\right)^{1/2}$, where $G$ and $\cal R$ are the
  gravitational and the gas constant. The mean Jeans mass $\langle
  M_{\rm J} \rangle$ is determined from average density in the
  system $\langle \rho \rangle$. An alternative cubic definition,
  $M_{\rm J} \equiv \rho (2\lambda_{\rm J})^3$, would yield a value
  roughly twice as large.}.
If scaled to mean densities $n({\rm H}_2) = 10^5\,$cm$^{-3}$, 
a value typical for star-forming molecular cloud regions
% (e.g.\ in $\rho$-Ophiuchus, see Motte, Andr{\'e}, \& Neri 1998)
and a temperature of 11.4$\,$K (i.e.\ a sound speed $c_{\rm s} =
0.2\,$km$\,$s$^{-1}$), the mean Jeans mass is one solar mass,
$\langle M_{\rm J} \rangle = 1\,$M$_{\odot}$, and the size of cube G
is $0.34\,$pc whereas all other models have a size of 0.29~pc.
The global free-fall time scale, as defined by $\tau_{\rm ff} =
(3\pi/32G)^{1/2}\,\langle\rho\rangle^{-1/2}$ with $\langle\rho\rangle$
being the average density, is about
$10^5\,$yr. In normalised time units it follows $\tau_{\rm ff}
= 1.5$. The  simulations cover a density range from $n({\rm H}_2) \approx
100\,$cm$^{-3}$ in the lowest density regions to $n({\rm H}_2)
\approx 10^9\,$cm$^{-3}$ where collapsing protostellar cores are
identified and converted into ``sink'' particles in the SPH code.

In this density regime gas cools very efficiently and it is possible
to use an effective polytropic equation-of-state in the simulations
instead of solving the detailed equations of radiative transfer. The
effective polytropic index is typically close to unity, $\gamma_{\rm
eff} \sil 1$, except for densities $10^5\,$cm$^{-3} < n({\rm H}_2) <
10^7\,$cm$^{-3}$, where somewhat smaller values of $\gamma_{\rm eff}$
are expected (Spaans \& Silk 2000). For simplicity, we adopt a value of
$\gamma_{\rm eff}=1$, i.e.\ an isothermal equation of state
for all densities in the simulations. As the choice of
$\gamma_{\rm eff}$ influences the density contrast in shock compressed
gas, this idealisation may introduce some small modifications to the
dynamical behaviour compared to real cloud systems (see Scalo et al.\
1998 or Ballesteros-Paredes et al.\ 1999b for further discussion).

\subsection{Model results}
Supersonic turbulence, even if strong enough to counterbalance gravity
on global scales will usually {\em provoke} local collapse.
Turbulence establishes a complex network of interacting shocks, where
converging shock fronts generate clumps of high density. This density
enhancement can be large enough for the fluctuations to become
gravitationally unstable and collapse.  Long-wavelength turbulence is
dominated by large-scale shocks which are very efficient in sweeping
up molecular cloud material, thus creating massive coherent
structures. When a coherent region reaches the critical density for
gravitational collapse its mass typically exceeds the local Jeans
limit by far.  Inside the shock compressed region, the velocity
dispersion is much smaller than in the ambient turbulent flow and the
situation is similar to localised turbulent decay.  Quickly a cluster
of protostellar cores forms. Therefore, all our models with zero
support, decaying, or large-scale driven turbulence, similarly lead to
a clustered mode of star formation. The efficiency of turbulent
fragmentation is reduced if the driving wavelength decreases. When
energy is inserted mainly on small spatial scales, the network of
interacting shocks is very tightly knit, and protostellar cores form
independently of each other at random locations throughout the cloud
and at random times with low efficiency corresponding to an isolated
mode of star formation (Klessen et al.\ 2000, 
Fig.\ 1 in Klessen 2001). The effect of magnetic fields strongly
depends on their strength. Whereas for the subcritical case, i.e. 
for fields which can support a gravitationally unstable region
magnetostatically (Mouschovias \& Spitzer 1976) no collapse occurs,
supercritical weak fields reduce the core growth rate perceptibly, 
although they cannot prevent local collapse (Heitsch et al.\ 2001).
%Supercritical magnetic fields are found 
%to play only a minor role in modifying the time scale and efficiency of local
%collapse.

To quantify the above we indicate in Tab.\ \ref{tab:models-2} the time
when local collapse has accumulated 5\% of the total mass in dense
cores and also give the core mass fraction after one free-fall
time scale. Hence, we can directly compare the $\Delta$-variance and
other the structural properties of the different models either at
equal times or at equivalent degrees of collapse.

\section{The evolution of structural properties}
\label{sec:results}

\subsection{The $\Delta$-variance analysis}

{\changed The $\Delta$-variance analysis was introduced by 
Stutzki et al.\ (1998) as an averaged wavelet transform 
(\cite{ziel99})
to measure the amount of structure present at different scales 
in an $E$-dimensional data set. The value of the $\Delta$-variance
at a certain scale is computed by convolving the $E$-dimensional
structure with a normalised spherically symmetric down-up-down 
function of the considered size, and measuring the remaining variance.
For two-dimensional structures, like astrophysical maps,
the filter function is easily visualized as a ``French hat'' wavelet 
with a positive centre surrounded by a negative ring and
equal diameters of the centre and the annulus. The analysis
can be applied in the same way in arbitrary dimensions by extending
the filter function to higher dimensions retaining its radial
symmetry and adapting the value in the negative part to preserve
normalization (Appendix B of Stutzki et al. 1998). }
The resulting $\Delta$-variance as a function of the filter size
measures the amount of structural variation on that scale.

{\changed As the convolution with the filter function corresponds
to a multiplication in Fourier space, we can relate the
$\Delta$-variance to the power spectrum of the data set.}
If the structure is characterised by a power law power spectrum
\begin{equation}
P(|\vec{k}|) \propto |\vec{k}|^{-\beta} 
\end{equation}
the slope $\alpha$ of the $\Delta$-variance as a function of 
lag (filter size) is related to the spectral index $\beta$ 
of the power spectrum by $\beta=\alpha+E$ for $0\le \beta < E+4$.
Due to the smooth circular filter function the $\Delta$-variance 
measures the spectral index in a way which is more robust with 
respect to edge and gridding effects than the 
Fourier transform. The $\Delta$-variance provides a clear spatial
separation of different effects influencing observed structures 
like noise or a finite observational resolution.

\mbox{}\cite{MLO} have shown that we can translate the
$\Delta$-variance of a three-dimensional isotropic structure into the
$\Delta$-variance of the maps obtained from the projections of this structure
by rescaling with a factor proportional to the lag and an
additional small shift.  This guarantees the preservation of the power
spectral index $\beta$ in projection (\cite{Stutzki}). As we want to
compare the simulations to observational data we will always use the
two-dimensional representation of the $\Delta$-variance results also
when applying the analysis directly to the three-dimensional data
cubes of the simulations. To compute the $\Delta$-variance for our
models the SPH density distribution is assigned onto a Cartesian grid
with $128^3$ cells.  The ZEUS cubes have been analysed in full
resolution at $256^3$ as well as degraded to $128^3$ for a one-to-one
comparison with the SPH models. {\changed Higher resolution helps to 
extend the dynamic scale range limited by the periodic boundary conditions
at the large scale end and the numerical resolution of the simulations
on the small scale end, but does not change the general behaviour
of} the resulting $\Delta$-variance.

\subsection{The evolution of the density structure}
\label{subsec:density}

\subsubsection{The collapse of a Gaussian density field}
\label{subsubsec:Gaussian-collapse}

Before we investigate the interplay between supersonic turbulence and
self-gravity, let us consider a system where the density and velocity
structure is dominated by gravity on all scales and at all
times. Model G describes the collapse of Gaussian density fluctuations
with initial power spectrum $P(k) \propto k^{-2}$ and maximum density
contrast $\delta \rho/\rho \approx 50$.  The system is unstable
against gravitational collapse on all scales and forms a cluster
of protostellar cores within about two free-fall time scales (Klessen \&
Burkert 2000, 2001). The velocity structure is coupled to the density
distribution via Poisson's equation and there is no contribution from
interstellar turbulence.

\begin{figure}[ht]
\unitlength1cm
\begin{picture}(8.6, 6.0)
\put( 8.6, 0.0){\begin{rotate}{90}\epsfysize=8.6cm \epsfbox{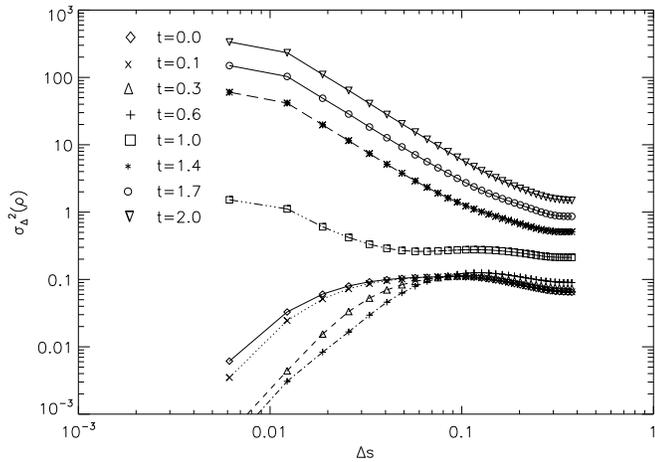}\end{rotate}}
\end{picture}
\caption{\label{fig:density-Gauss}Time evolution of the strength
of density fluctuations as function of their spatial scale  
measured with the $\Delta$-variance for model G. The density $\rho$
is given in units of the average density in the cube, the lag 
$\Delta s$ in units of the cube size, and the time $t$ in internal
time units where the free-fall time $\tau\sub{ff}=1.5$.
} 
\end{figure}

The time evolution of the density structure is illustrated in Figure
\ref{fig:density-Gauss}. Initially, the $\Delta$-variance
$\sigma_{\Delta}^2(n)$ is more or less constant ($\alpha=0$)
on scales $\Delta s \sig 0.02$, in agreement with the initial power 
spectrum $P(k)\propto k^{-2}$. The steepening below  $\Delta s \approx 0.02$
is produced by the finite resolution of the SPH simulations
resulting in the blurring of structures at the smallest scales.

The first changes of the variance $\sigma_{\Delta}^2(n)$ are confined
to small scales.  Initial fluctuations with masses below the
local Jeans limit will quickly smear out by thermal pressure as the
system evolves from purely Gaussian fluctuations into a
hydrodynamically self-consistent state (see Appendix B in Klessen \&
Burkert 2000).  As these fluctuations are by far more numerous than
Jean-unstable contracting ones, the $\Delta$-variance
$\sigma_{\Delta}^2(n)$ begins to decrease on small scales. However, as
the central regions of massive Jeans-unstable fluctuations contract to
sufficiently high densities, $\sigma_{\Delta}^2(n)$ increases again.
This mainly affects the small scales as local collapse modifies the density
structure on time scales of the local free-fall time. At
$t=0.7\tau\sub{ff}$ the first collapsed core is identified and is soon
followed by others.  Altogether 56 dense protostellar cores build up.
As time advances larger and larger scales exhibit noticeable signs of
contraction.  After about one global free-fall time
collapse starts to involve all spatial modes in the system and
the absolute magnitude of the density fluctuations finally grows on
all scales.  As the small scale structure dominates the density
structure we obtain a negative slope in the $\Delta$-variance
spectrum. In the final step of the simulation roughly 30\,\% of the mass is
accumulated in dense cores and the slope is about --1.7 indicating that a
small but significant contribution of large scale structure is still
present, because an uncorrelated $N$-body system of gravitationally
collapsed points would correspond to a slope of --2 equivalent to a
flat power spectrum $P(k)={\rm const}$.

The flattening at $\Delta s > 0.2$ is due to periodicity. The system
is {\em not} allowed to collapse freely, it is {\changed held} up against global
collapse by periodic boundaries which strongly affect the evolution of the
large-scale modes. The graphs of $\Delta$-variance are not extended beyond
effective lags of about 0.4 as the largest filter that we use is half the
cube size and we have to apply an average length reduction factor of $\pi/4$ on
projection to two dimensions.

\subsubsection{The interaction between gravitation and turbulence}
\label{subsubsec:HD-turb}

\begin{figure*}[ht]
\unitlength1cm
\begin{picture}(17.6,12.5)
\put(  8.6, 6.5){\begin{rotate}{90}\epsfysize=8.6cm \epsfbox{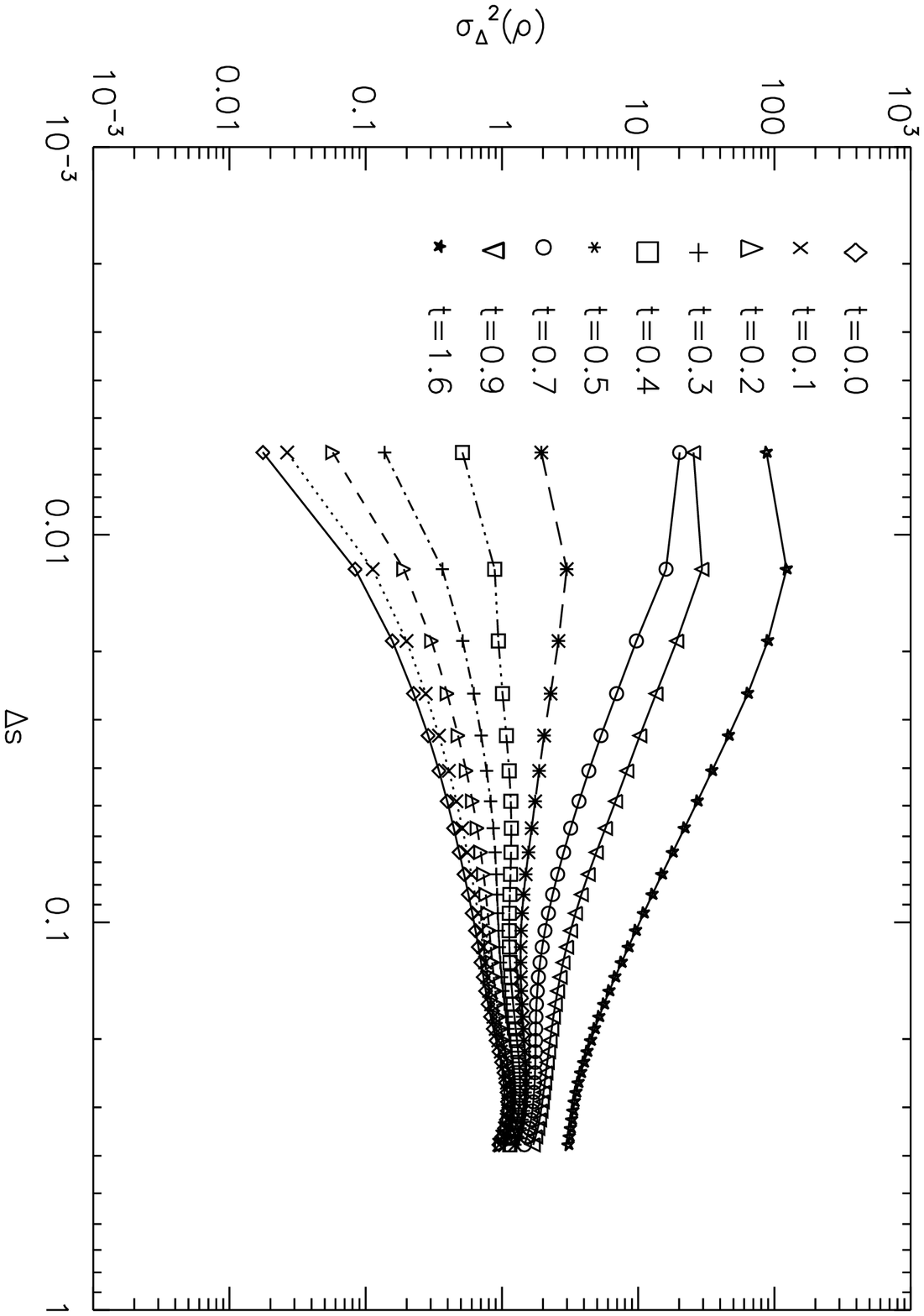}\end{rotate}}
\put( 17.6, 6.5){\begin{rotate}{90}\epsfysize=8.6cm \epsfbox{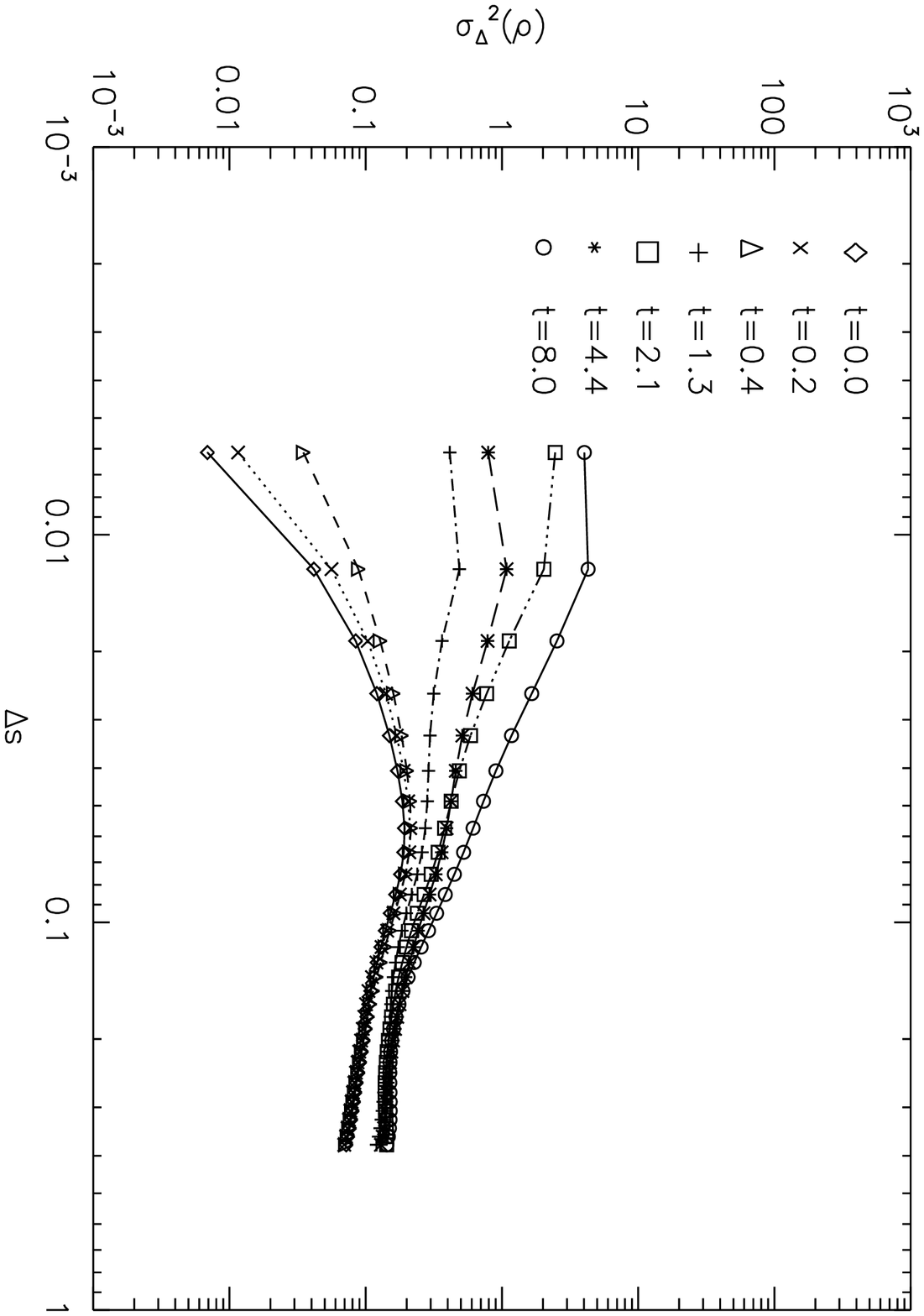}\end{rotate}}
\put(  8.6, 0.0){\begin{rotate}{90}\epsfysize=8.6cm \epsfbox{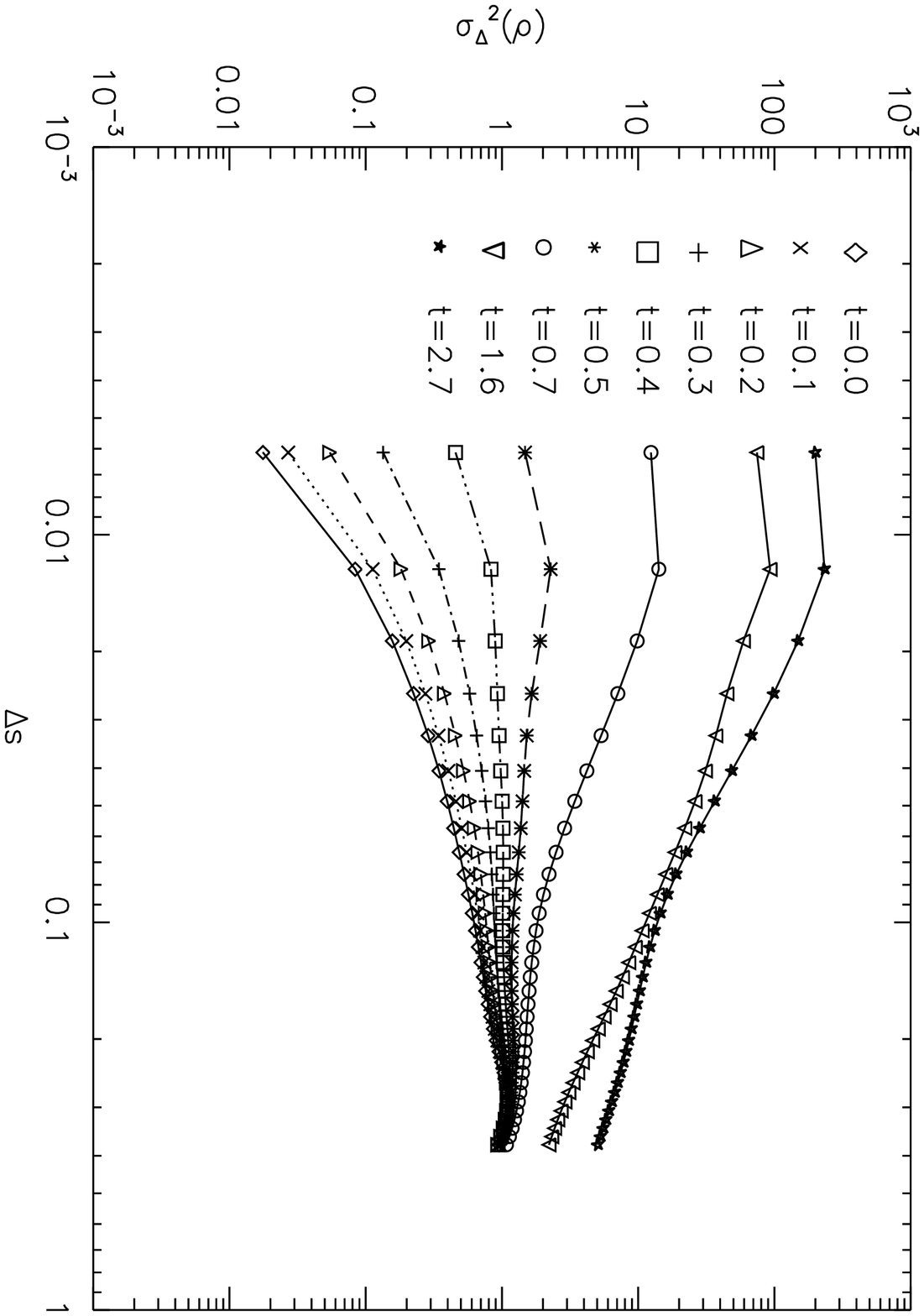}\end{rotate}}
\put( 17.6, 0.0){\begin{rotate}{90}\epsfysize=8.6cm \epsfbox{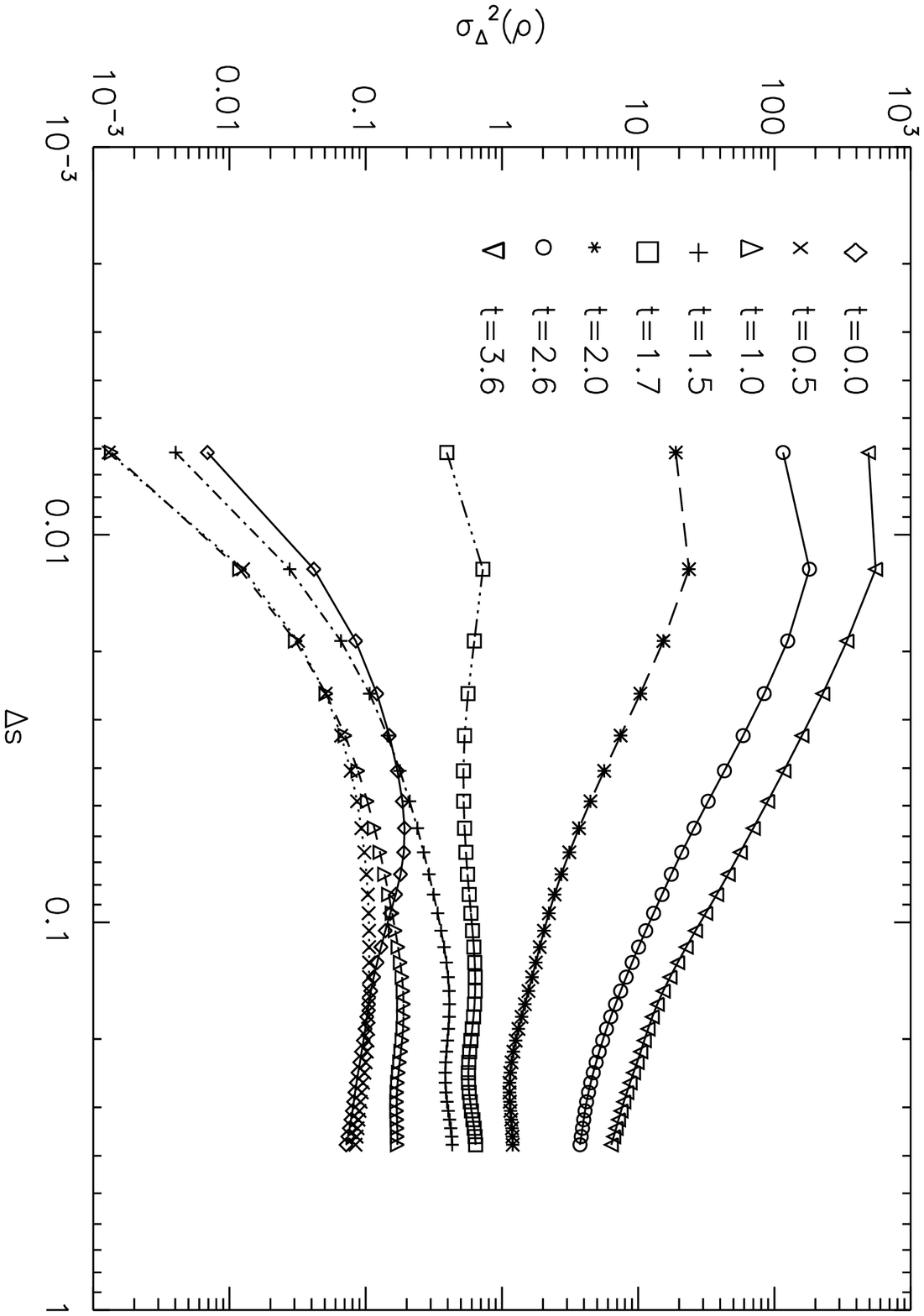}\end{rotate}}
\put(  7.9,11.8){\bf a)}
\put( 16.9,11.8){\bf b)}
\put(  7.9, 5.3){\bf c)}
\put( 16.9, 5.3){\bf d)}
\end{picture}
\caption{\label{fig:density-turb}Time evolution of the size
distribution of density variations as measured with the 
$\Delta$-variance for the four models of (a) S01, (b) S02, (c) Sd1, and (d) Sd2.
The different depicted times are indicated at the
left side of each plot.}
\end{figure*}

To study the interplay between supersonic turbulence and self-gravity,
we consider four models of interstellar turbulence which probe
very disparate regions of the relevant parameter space. In models S01 and
Sd1 most of the turbulent kinetic energy is carried on large scales,
whereas models S02 and Sd2 involve mainly small-scale turbulent modes. 
In S01 and S02, turbulence is continuously driven such that at any 
moment the overall kinetic energy compensates the global gravitational energy.
In Sd1 and Sd2, the turbulent energy is allowed to decay freely. 

Figure \ref{fig:density-turb} shows $\sigma_{\Delta}^2(n)$ for
all four models as function of time.  In the initial plots one
can clearly see the dominance of the driving scale as discussed
by \cite{MLO}. The introduction of a velocity field with a
certain scale induces a pronounced peak in the density structure
at a somewhat smaller scale. Thus the curves at $t=0$ show
for the large scale-driven models a power law $\Delta$-variance
from about a third of the cube size down to the numerical
dissipation scale whereas in the small-scale driven model the
driving feature at about 0.07 dominates the structure.

Star formation is a joint feature of all considered models.  Like in the
evolution of the Gaussian density field the gravitational collapse
first modifies only the smallest scales, hardly changing the global
behaviour.  As soon as local collapse occurs and the first dense
protostellar cores form and grow in mass by accretion, they represent
the main density fluctuations. Their power is concentrated on small scales 
and the $\Delta$-variance exhibits a negative slope. The structure resulting
from the collapse is very similar in the various models when we
compare evolutionary stages with about the same mass fraction
collapsed onto cores.

In the initially large-scale driven models a turbulent cascade covering
all scales is already present from the beginning shown by the highest 
$\sigma_{\Delta}^2(n)$-values at long  scales. Within this cascade the 
number of small-scale fluctuations is small compared to models S02, Sd2,
and G, and they are typically part of a larger structure so that they
are only weakly dispersed in time. This leads  to a monotonous growth
of the $\Delta$-variance on all scales.
As larger and larger regions become gravitationally unstable the
contraction comprises increasingly larger scales, and finally
$\sigma_{\Delta}^2(n)$ shifts ``upward'' on all scales while
maintaining a fixed slope.  This situation is similar in {\em all}
models with allow for large-scale collapse, i.e.  it is only prevented
in the small scale-driven model S02.

A different temporal behaviour is visible in Sd2, the decaying turbulence
which was originally driven at small scales. Like in the collapse of
the Gaussian density field we find in the first steps of the
gravitational evolution a relative reduction of small-scale structure.
This decrease is due to the termination of the initial small-scale
driving resulting in a quick dissipation of the existing fluctuations
by thermal pressure if they are not Jeans supercritical, analogously
to the Gaussian collapse case. In the next steps we notice the
production of structure on larger scales. The smoothing of small-scale
turbulence combined with the onset of self-gravity leads to global
streaming motions which produce density structures correlated on a larger
scale. {\changed Large-scale structures had been initially} suppressed 
by the non-local turbulent driving
mechanism. After less than one free-fall time a kind of self-sustaining
inertial cascade with a $\Delta$-variance slope $\alpha$ of about 0.5
is build up like in all other decaying models and in the large scale
driven model. After these initial adjustments the first protostellar
cores form and we find the same dynamical behaviour.

The time scale to reach a comparable collapse state and the final 
structure that we reach in the simulations differs
between the models, mainly determined by the strength of the 
turbulent driving. The exponent
$\alpha$ of the $\Delta$-variance in the collapsed state is --1.3 in
the models containing continuous driving, --1.5 in the models where
the turbulence decays during the gravitational collapse, and --1.7 for
the pure collapse of the Gaussian density field. This is
understandable, as decreasing turbulent support leads to enhanced
collapse forming stars in denser clusters. The final deviation from
the uncorrelated field of protostars which has $\alpha=-2$ is thus a
measurable indicator of the turbulent processes in the cloud during
gravitational collapse.  The local collapse produces ``point-like''
high-density cores with small overall filling factor whereas most
of the volume is supported by hydrodynamic turbulence.
In the large-scale driven model 30\,\% of the mass and in the decaying
models 60\,\% of the total mass has turned into cores within the considered 
time interval. In the model
continuously driven at small scales only 3\,\% of the mass is in
stable cores despite a considerably longer simulation time.  The
small-scale driven turbulence leads to the least efficient star
formation in an isolated mode, whereas the other cases result in the
formation of stellar aggregates and clusters (see also Klessen et al.\
2000, Klessen 2001).

Beside the deceleration of collapse the continuation of 
the initial driving has almost no influence on the 
resulting density structure as soon as the first stable 
cores have formed.
It only {\changed maintains} a constant level of velocity fluctuations
in the main volume of the cloud which is dominated by
low density gas, compared to a homogeneous reduction of these 
fluctuations in the purely decaying case.

\subsubsection{The influence of the numerical model}
\label{subsubsec:SPH-ZEUS-comparison}

\begin{figure}[ht]
\unitlength1cm
\begin{picture}(8.6,12.5)
\put(  8.6, 6.5){\begin{rotate}{90}\epsfysize=8.6cm \epsfbox{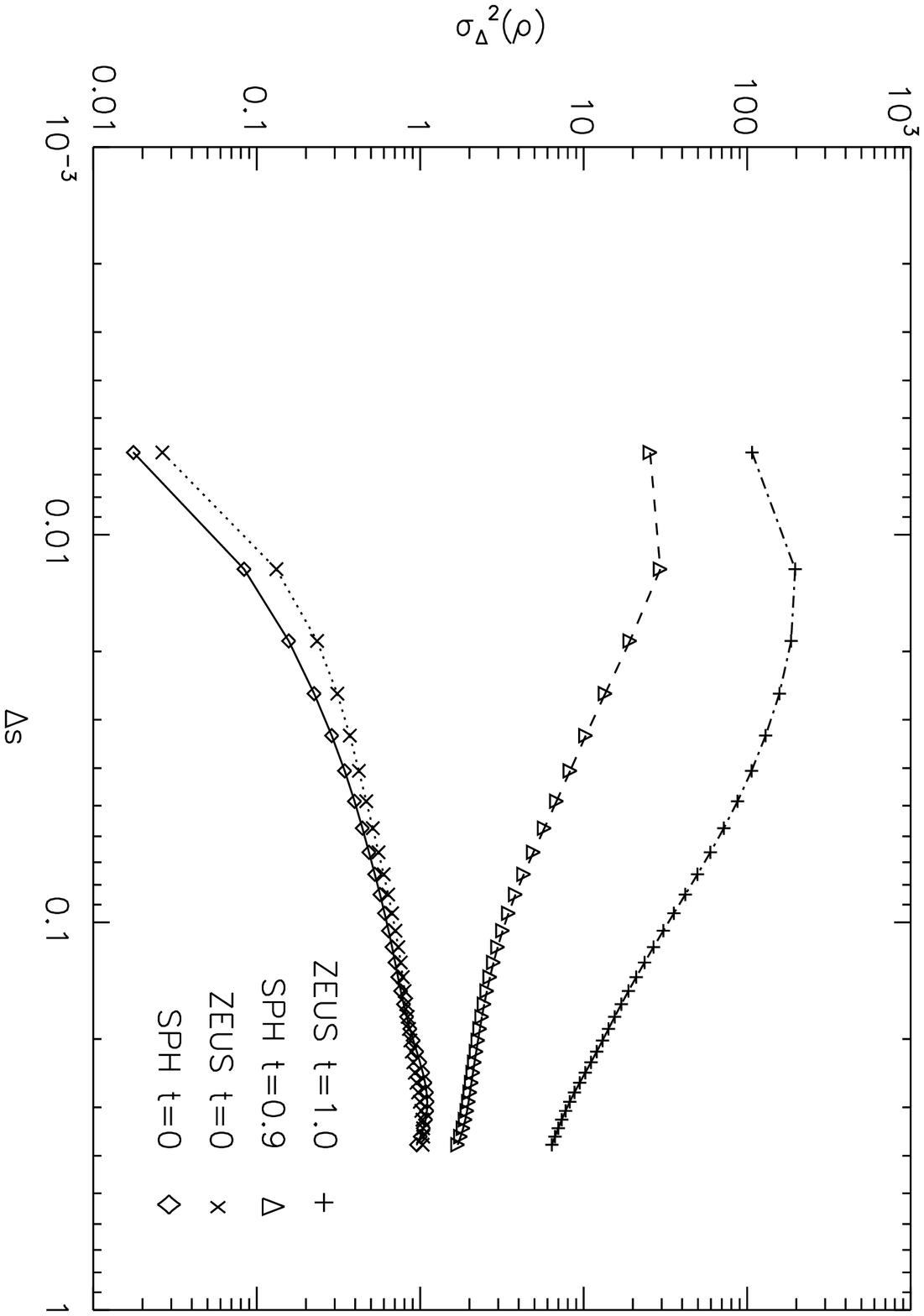}\end{rotate}}
\put(  8.6, 0.0){\begin{rotate}{90}\epsfysize=8.6cm \epsfbox{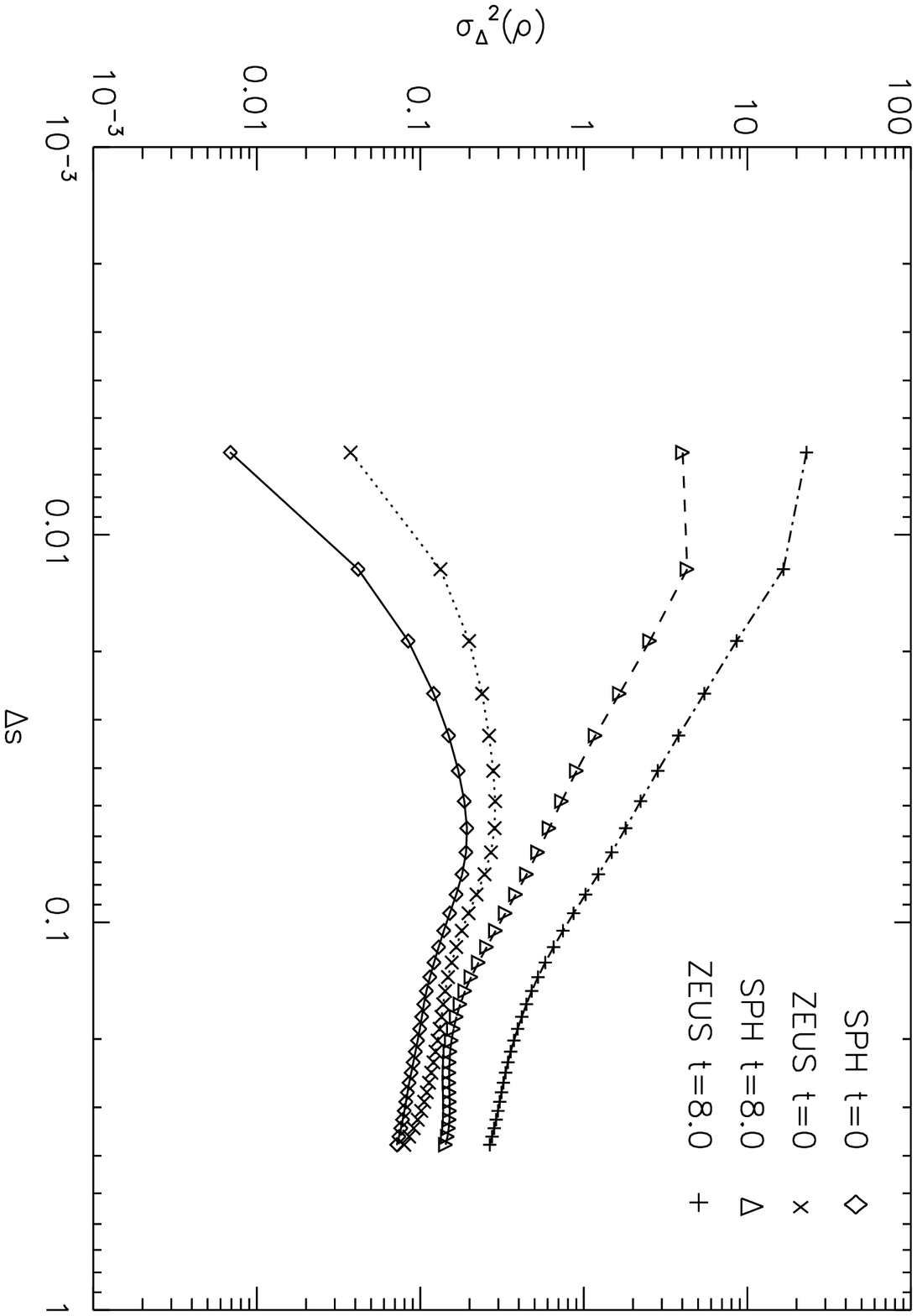}\end{rotate}}
\put( 1.3, 11.8){\bf a)}
\put( 1.3, 5.3){\bf b)}
\end{picture}
\caption{\label{fig:SPH-ZEUS}Comparison of the $\Delta$-variance
measured from the particle-based and the grid-based simulations
at the beginning and at about the same timestep of the gravitationally
collapsed state. Plot (a) shows the large-scale driven model,
plot (b) the small scale-driven situation.}
\end{figure}

Comparing the results of the particle-based SPH and the grid-based
ZEUS code we can distinguish between numerical artifacts and physical
results, as these two approaches practically bracket the real
dynamical behaviour of interstellar turbulence.
In Fig.\ \ref{fig:SPH-ZEUS} we compare the $\Delta$-variance plots
of the density structure obtained for the driven cases using 
either SPH or ZEUS, at the beginning of the gravitational collapse and
in a step where the structure is already dominated by protostellar
cores.

The scaling behaviour of the density structure does not differ between
both types of simulations but the absolute magnitude of the density
fluctuations as seen in the total value of the $\Delta$-variance is
somewhat larger for all ZEUS models. In the first steps of the
large-scale driven models both numerical approaches still agree
approximately but during the evolution the scale dependent density
variations become about a factor five higher in the ZEUS model than in
the SPH approach. In the small scale driven models we can notice a
clear difference already at the beginning of the simulations. This is
consistent with the different effective resolution of the methods.
Whereas the SPH code can provide a very good spatial resolution around
the collapsing dense regions the general spatial resolution obtained
with $2\,10^5$ particles is lower than in the ZEUS simulations on a
$256^3$ grid. Thus, the damping of structures at small scales 
due to the finite resolution of the code is slightly stronger in the
SPH simulations than in the corresponding ZEUS models. 
One can for instance see that there is a virtual
reduction of structure below 0.01 which is approximately the radius of
the sink particles in the SPH code. 

As the SPH resolution is explicitly density dependent it is also
reduced on all larger scales in low density regions. This virtually smears
out part of the structure on all scales. Consequently, the $\Delta$-variance
shows lower values on all scales than in the grid-based approach because
it is not biased towards high-density regions like SPH and the observations.
The effect is larger in the small-scale driven models as the same number
of SPH particles has to {\changed represent more shocks than in
large-scale dominated cases} further reducing the effective resolution. 
Moreover, the resolution worsens during the collapse evolution
as SPH particles ``vanish'' in the sink particles. Thus, the
ZEUS simulations are preferential due to their higher resolution
if one is interested in the absolute value of the $\Delta$-variance
whereas they provide no essential advantage for the study of the
scaling behaviour.

\subsubsection{Magnetic fields}
\label{subsubsec:MHD-turb}

\begin{figure}[ht]
\unitlength1cm
\begin{picture}(8.6, 6.0)
\put( 8.6, 0.0){\begin{rotate}{90}\epsfysize=8.6cm \epsfbox{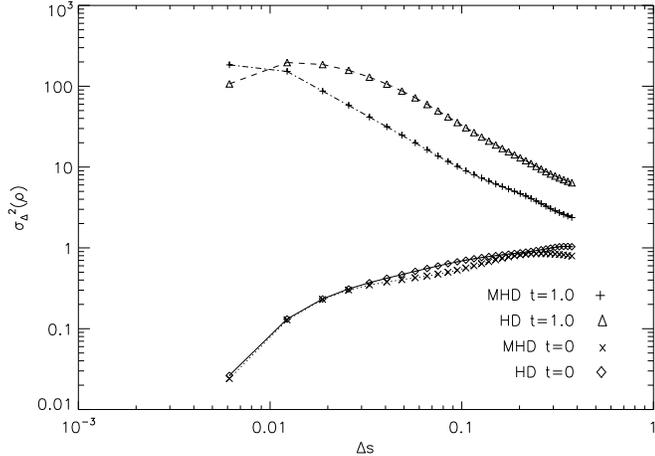}\end{rotate}}
\end{picture}
\caption{\label{fig:density-MHD}Comparison of the $\Delta$-variance
for the large-scale driven model in the hydrodynamic case
or the situation with strong magnetic fields, both computed
with the ZEUS code.}
\end{figure}

As discussed by \cite{OML} magnetic fields hardly change the
general scaling behaviour in interstellar turbulence but create
anisotropies in the velocity field and therefore aligned density
structures. Since the $\Delta$-variance cannot measure anisotropies
in the density structure we do not expect to detect the influence
of the magnetic field on the turbulence by the present analysis. 
Fig. \ref{fig:density-MHD} shows the $\Delta$-variance
for the initial step and an collapsed stage in a large-scale
driven hydrodynamic model and the equivalent MHD model with a 
strong magnetic field. The initial steps are almost identical but
we find that during collapse the magnetic field effectively helps to transfer 
structure from larger to smaller scales. Thus we confirm the more 
qualitative conclusion of Heitsch et al.~(2001) that the magnetic
field slightly delays collapse by transferring part of the
turbulent kinetic energy to smaller scales. The general slope
of the $\Delta$-variance is not changed but we obtain somewhat
denser and smaller cores and somewhat less large-scale correlation at
equivalent timesteps.

Computing the $\Delta$-variance for maps projected either in the
direction of the initial magnetic field or perpendicular to it does
not show any significant difference in the density scaling 
behaviour as mainly the shape of the collapsed regions is influenced,
towards spiral-shaped structures, which is not measurable with
the isotropic $\Delta$-variance filter.

We have also tested models with {\changed a smaller} magnetic field where the
magnetic pressure in the order of the thermal pressure or lower.
Here, we find that the field acts like an additional 
contribution to the overall isotropic pressure so that the collapse
is somewhat delayed relative to the hydrodynamic case but the
general structure does not deviate from the hydrodynamic simulations.
Thus there is no need to discuss the weak-field situation here
separately.

\subsection{The velocity structure}
\label{subsec:velocity}

\begin{figure}[ht]
\unitlength1cm
\begin{picture}( 8.6, 6.0)
\put( 8.6, 0.0){\begin{rotate}{90}\epsfysize=8.6cm \epsfbox{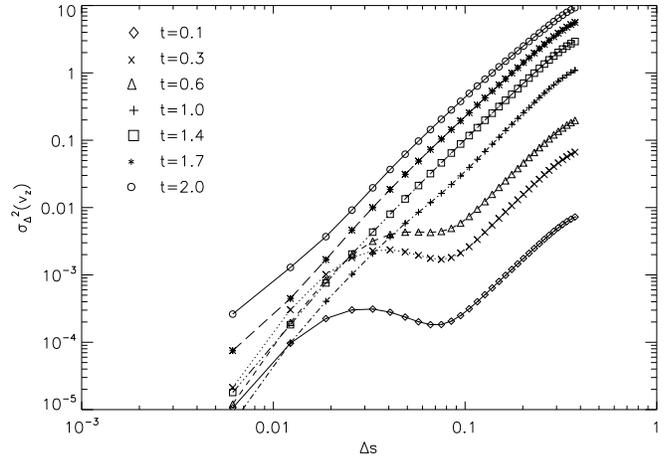}\end{rotate}}
\end{picture}
\caption{\label{fig:velocity-gaussian}Evolution of the $\Delta$-variance of 
the $z$-velocity in the collapse of the Gaussian density distribution
(model G).  The velocity $v_z$ is given here in units of the 
thermal sound speed $c\sub{s}$.}
\end{figure}

We can apply the $\Delta$-variance analysis in the same way to the velocity
structure in the simulations (Ossenkopf \& Mac Low 2001).
Fig. \ref{fig:velocity-gaussian} shows the evolution of one velocity
component in the collapse of the Gaussian density fluctuations. In the
first steps where we observe a relative reduction of small scale density
fluctuations we find a bimodal velocity distribution with either very
small or very large flows. The surplus of small-scale flows just reflects
the dissipation of the initial small-scale variations by thermal pressure.
When the first stable cores have formed
the picture changes towards that of a typical shock-dominated medium
with a slope $\alpha=2$ (Ossenkopf \& Mac Low 2001) as the result
of supersonic accretion onto dense cores along the emerging filamentary
structure (Klessen \& Burkert 2000, 2001).

\begin{figure}[ht]
\unitlength1cm
\begin{picture}( 8.6, 6.0)
\put( 8.6, 0.0){\begin{rotate}{90}\epsfysize=8.6cm \epsfbox{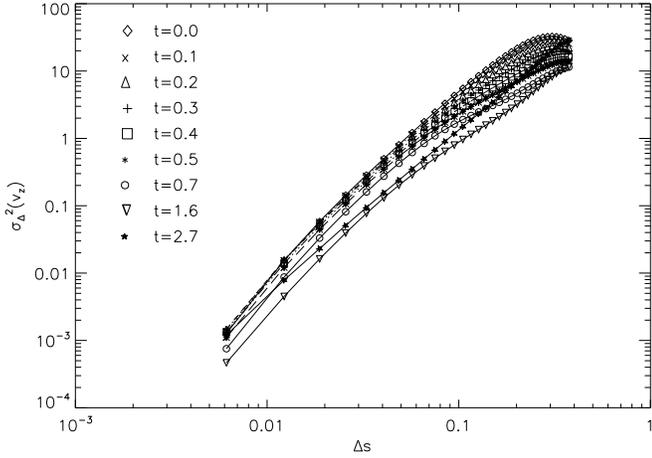}\end{rotate}}
\end{picture}
\caption{\label{fig:velocity-decaying}Evolution of the
$\Delta$-variance of the $z$-velocity component of the decaying
model Sd1.
}
\end{figure}

{\changed %RSK: habe den Paragraph etwas umgeschrieben...  
In all driven models we see no significant changes in the velocity
structure during collapse. This is because the $\Delta$-variance is
not focused towards the dense cores where collapse motions occur, as
these have only a small spatial filling factor. Instead, most of the
volume is occupied by tenuous intercore gas with velocity structure
that is determined by turbulent driving. The $\Delta$-variance
therefore exhibits the power-law behavior of shock dominated gas.

The same hold for decaying turbulence as well. To illustrate that
point, Fig. \ref{fig:velocity-decaying} presents the evolution of the
velocity structure for model Sd1, where we drive the turbulence
initially at large scales and switch off the driving during the
gravitational collapse.  The changes in the velocity structure are
only minute. The $\Delta$-variance follows the power law of
shock-dominated flows throughout the entire evolution. Only the total
magnitude of the velocity fluctuations decreases slightly during the
initial decay of turbulence. However, after the onset of collapse,
when the majority of mass is already accumulated in dense cores, the
magnitude of $\sigma_{\Delta}$ increases again. The $\Delta$-variance
becomes dominated by the shock structure arising from the supersonic
accretion flows onto individual cores (similar to the late stages of
model G). The evolution of the small-scale decaying turbulence is not
plotted separately, as we observe the same behavior. Except during the
initial phase of turbulent decay where the velocity structure still
peaks on small scales reflecting the smaller driving wavelength used
to set up the model (see Fig.\ \ref{fig:SPH-ZEUS}d for the density
structure), again after $t\approx 1.5$ when the initial turbulence
is sufficiently decayed away the $\Delta$-variance arrives at the power-law
behavior of shocked gas.  }

{\changed As a point of caution, we note here that} the velocity
structure of all
SPH models exhibits a numerical artifact due to the density weighted
selection of the size used for the averaging kernel. This produces
virtual velocity streams with relatively large coherence length at low
densities resulting in a $\Delta$-variance slope $\alpha$ of about 2.8
significantly above the value of 2 characteristic for a
shock-dominated medium. As the weighting function is given by the
density of SPH particles, this effect is only visible in the velocity
structure and not in the density scaling.  The corresponding
grid-based models do not exhibit a similar enhancement of large scales
in the velocity structure and show a $\Delta$-variance slope of about
2.

\begin{figure}[ht]
\unitlength1cm
\begin{picture}( 8.6, 6.0)
\put( 8.6, 0.0){\begin{rotate}{90}\epsfysize=8.6cm \epsfbox{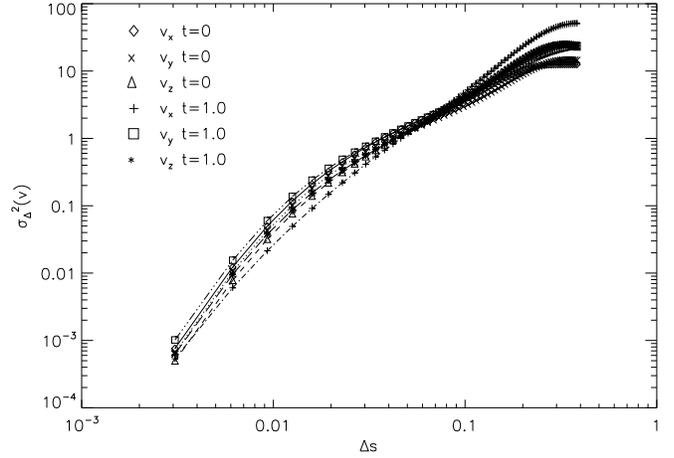}\end{rotate}}
\end{picture}
\caption{\label{fig:velocity-mhd}$\Delta$-variance of all
three velocity components for model the large-scale driven
MHD model at the initial step and after one free-falling time.
}
\end{figure}

In Fig. \ref{fig:velocity-mhd} we show the three velocity components
of the MHD model M01 at the same two timesteps like in the density plot in 
Fig. \ref{fig:density-MHD}. In contrast to the findings of
\cite{OML} for sub-Alfv\'{e}nic turbulence, we see no strong 
anisotropy of the velocity field. The velocity structure
along the mean magnetic field ($z$-direction) is very similar to the
perpendicular directions throughout the dynamical
evolution of the system and well within the statistical fluctuations
expected for large-scale turbulence. The velocity structure is still
determined by supersonic turbulence rather than by the magnetic field 
structure because the turbulent rms velocity dispersion exceeds the 
Alfv\'{e}n speed in this model. The local collapse
and the formation of a cluster of collapsed cores tends to make the
influence of the magnetic field on the velocity structure even weaker.
The magnetic field merely decelerates the gravitational
collapse  and changes the geometry of the collapsing regions
as seen in Fig.\ \ref{fig:density-MHD}, but hardly
changes the global velocity structure in this model situation.

Altogether we find that all models that are allowed to evolve freely
or are driven at large scales exhibit a similar velocity scaling
behaviour, characteristic of shock-dominated media.  This is the effect of
the undisturbed turbulence evolution and the appearance of
accretion shocks. Both effects lead to remarkably
similar properties of the velocity $\Delta$-variance. Any observed
deviation from this large-range ``Larson'' behaviour
indicative of shock-dominated media will hint the presence of
additional physical phenomena and could provide constraints on the
initial conditions and the dynamical state of star-forming regions.

\section{Comparison with Observations}
\label{sec:observations}

\subsection{Dust observations}

To compare the simulations of gravitational collapse with
observational data of collapsed regions we have selected the observation
of a star-forming cluster in Serpens by \cite{Testi}
as this region represents a state of star formation that may be 
similar to the outcome of our simulations.

\begin{figure}[ht]
\unitlength1cm
\begin{picture}(8.6, 6.0)
\put( 8.6, 0.0){\begin{rotate}{90}\epsfysize=8.6cm
\epsfbox{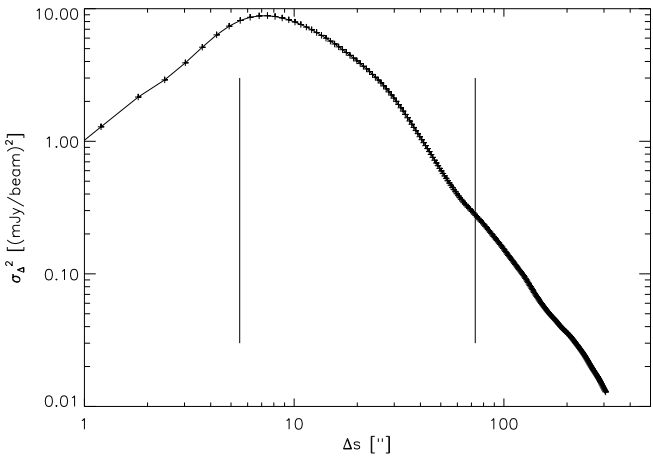}\end{rotate}}
\end{picture}
\caption{\label{fig:testi}$\Delta$-variance of the dust continuum
map in Serpens taken by \cite{Testi}. The two vertical lines represent
the limits of the significant range as indicated with the
observational data.
} 
\end{figure}

In Fig. \ref{fig:testi} we show the $\Delta$-variance for the 3~mm
dust continuum map of \cite{Testi}. It reveals an increase of the
relative amount of structure from small scales towards a peak at
7$''$, an intermediate range which can be fitted by a power-law
exponent $\alpha=-1.2$, and a decay with $\alpha=-2$ indicating the
complete lack of large-scale structure at lags above 40$''$. The
behaviour at largest and smallest scales can be understood when
looking at the observational base of the map. \cite{Testi} give a
resolution for their interferometric observations of $5.5'' \times
4.3''$. Consequently we cannot see any structure below that size. The
fact that our peak falls with 7$''$ somewhat above the 5.5$''$
resolution limit might indicate that the CLEAN beam used in the reduction of
the interferometric data is not exactly Gaussian or slightly wider
than computed. The whole map is taken with an interferometric
mosaicing technique (see \cite{Testi1} for details) without a 
zero-spacing by complementary single dish observations.  Thus the map
cannot contain any structure on scales above the single pointing areas
determined approximately by the size of the primary beam of the OVRO
antennas of 73$''$. This is in agreement with the lack of structure
indicated by the $\Delta$-variance slope of --2 at these scales. 
The two limiting sizes are indicated by vertical lines in
Fig. \ref{fig:testi}. Thus we may only discuss the range in between
disregarding other information that is plotted in the interferometric
map but that can eventually not be obtained from the observations.

The steepening of the $\Delta$-variance in the intermediate size
range from $\alpha=-1.2$ to $\alpha=-2$  does qualitatively
agree with the behaviour observed in most collapse simulations 
at small scales but does not match any of them quantitatively.
For a detailed comparison the dynamic scale range covered in the simulations
is still insufficient due to the periodic boundary conditions
constraining the large scale behaviour. Hence, we can only
conclude that the collapse models show the same general structure
as the dust observations, indicating that they represent a 
realistic scenario but we cannot yet discriminate between different
models using the observational data.

\subsection{Molecular line observations}

\mbox{}\cite{Bensch} provided a detailed $\Delta$-variance analysis 
of the density structure traced by observations in different CO isotopes
for several molecular clouds with different states of
star formation including quiescent clouds like the Polaris Flare
and clouds with violent star formation like Orion A.
They found for all molecular clouds a density structure approximately
characterised by a power law $\Delta$-variance, with an exponent in the
range $0.5 \le \alpha \le 1.3$.  In the best studied cloud one smooth curve
connects scales larger than 10$\,$pc (where turbulence presumably is
driven) with the dissipation scale at 0.05$\,$pc (where ambipolar
diffusion processes become important). The positive slopes 
indicate that the density structure seen in the CO isotopes is dominated by
large-scale modes. This result is consistent with purely supersonic 
turbulence and appears independent of the dynamical
state of the molecular cloud regions studied, i.e. regardless
whether the cloud forms stars or not. 

This is somewhat surprising, since we expect that the density distribution 
in star-forming regions is dominated by the collapsing protostellar cores 
on small spatial scales.
The $\Delta$-variance spectrum therefore should exhibit a {\em negative} 
slope as we demonstrate in \S\ref{sec:results}. 
The molecular line results are also in obvious contradiction to the 
Serpens dust observations discussed above.

The explanation for the difference is hidden in the radiative
transfer problem. A discussion of all major aspects of molecular
line transfer in turbulent media is beyond the scope of this study
and will be provided in a separate paper (Ossenkopf in prep.).
Here, it is sufficient to concentrate on one effect
-- saturation at large optical depths. Molecular lines like the 
lower transitions of ${}^{13}$CO, frequently used to map the
density profile of molecular clouds, become typically optically thick
in the cores of clouds at densities in the order of $10^5$ cm$^{-3}$.
The exact value depends
on the transition, the spatial configuration, temperatures,
and the geometry of the radiation field but one can always assign
a typical density range to the transition from the optically thin
to the optically thick regime. This leads to a saturation of the
line intensities in dense clumps so that the lines do not trace their
internal structure but rather see clump surfaces.
Moreover, the molecules tend to freeze out in dense dark
regions (Kramer et al.\ 1999) amplifying the effect that the line brightness
reflects only part of the column density in dense clumps.

As we do not want to treat the full radiative transfer problem
here, we give only an estimate for the influence of optical
depth effects by including a saturation limit into our
computations. Because the simulations are scale-free and the
typical saturation density varies for different molecules and transitions
there is no particular density value to be used for this limit
so that we have to play with different values.

\begin{figure}[ht]
\unitlength1cm
\begin{picture}(8.6,12.5)
\put(  8.6, 6.5){\begin{rotate}{90}\epsfysize=8.6cm \epsfbox{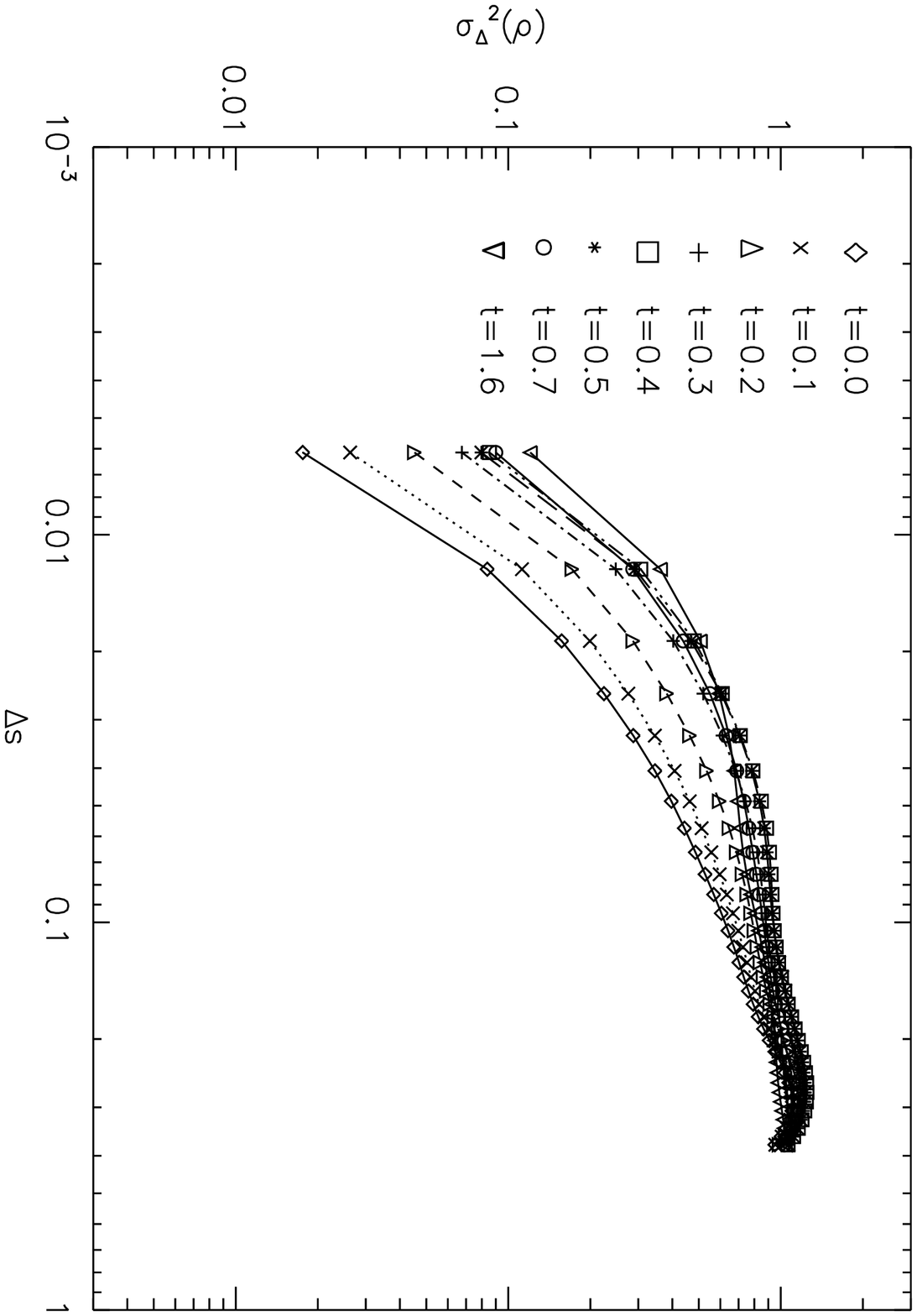}\end{rotate}}
\put(  8.6, 0.0){\begin{rotate}{90}\epsfysize=8.6cm \epsfbox{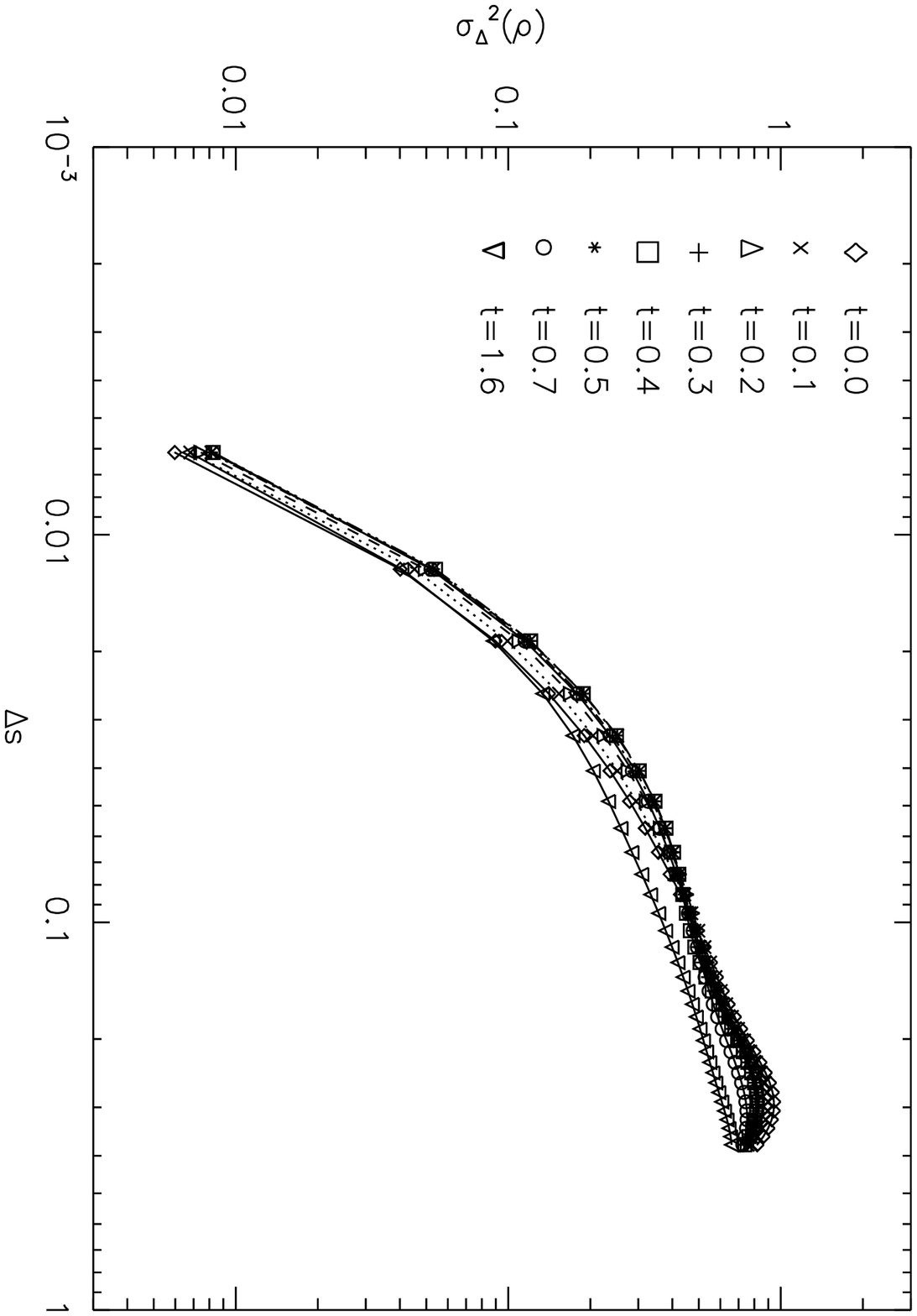}\end{rotate}}
\put( 7.9, 11.8){\bf a)}
\put( 7.9, 5.3){\bf b)}
\end{picture}
\caption{\label{fig:cutted}Time evolution of the $\Delta$-variance for 
the large-scale driven model S01 when the density structure is assumed to
saturate at densities of 240.0 (a) and 24.0 (b).}
\end{figure}

Figure \ref{fig:cutted} illustrates the evolution of the large-scale
driven model shown in Fig. \ref{fig:density-turb}a assuming now 
that all densities above a certain threshold are invisible so that 
they are equal to the value of the saturation density. In the upper plot we
have chosen this limit to be the maximum density occurring in the
original turbulent density distribution before gravitational collapse
starts. This is 240 times the average density, i.e.\ a relatively
large value compared to the dynamic range of molecular line observations.
In the lower graph the saturation limit is reduced by a factor
10.

Although only a relatively small fraction of the material appears
at densities above the limit the influence on the $\Delta$-variance is
dramatic. As the collapsing cores produce the relative
enhancement of small scale structure their virtual removal by the saturation
results in an almost constant $\Delta$-variance behaviour during the
gravitational collapse. Even in the very conservative upper plot where we
assume that all structures occurring in normal interstellar turbulence
are still optically thin and only the collapsing cores become
optically thick the $\Delta$-variance stays at a positive slope during
the entire evolution.  In the other case where optical depth effects
are assumed to be more important the $\Delta$-variance remains at a
fixed slope of $0.5$ consistent with the molecular line observations
of interstellar clouds.

Hence, optical depth effects can easily prevent the detection of 
gravitational collapse in molecular line observations, since they reproduce
the $\Delta$-variance spectrum of a turbulent molecular cloud even if
collapse has lead already to the formation of protostellar cores.
Even in massively star-forming clouds most molecules used for mapping
trace the diffuse density structure {\em between} Jeans-unstable 
collapsing protostellar cores. This gas is still dominated by interstellar
turbulence. The density contrast in star-forming molecular cloud regions simply
exceeds the density range traceable by molecular transitions.
Protostellar core densities are so high that ${}^{13}$CO at best traces the
outer envelope. Therefore both, star-forming and quiescent molecular 
clouds, exhibit very similar molecular line maps.

It is essential to resolve large density contrasts measuring
the full density structure to study the influence of self-gravity and 
local collapse in star-forming clouds. This can be
achieved using dust continuum emission. Indeed, the 3mm continuum map
of the Serpens cluster by Testi \& Sargent (1998) shows a density
structure that is dominated by small scales as predicted by
our collapse simulations. The drawback
of dust emission observations is the inherent convolution of the
density structure with the unknown temperature profile. Large
dust extinction maps could circumvent this problem but require
long integration times at NIR wavelengths to obtain a dense
sampling with background stars (Lada et al.\ 1999).

\section{Summary}
\label{sec:summary}

In this paper we investigate the time evolution of
the structural characteristics of self-gravitating supersonically
turbulent gaseous systems using the $\Delta$-variance as
statistical measure. We resolve the transition from purely
hydrodynamic turbulence to gravitational collapse, 
the formation and mass growth of protostellar cores, and compare
different models of driven and freely decaying self-gravitating
turbulence with and without magnetic fields.

Contrary to what is observed for purely hydrodynamic turbulence,
self-gravitating supersonic turbulence yields a density structure that
contains most power on the smallest scales (i.e. in the collapsed
objects) as soon as local collapse has set in. This happens
in all self-gravitating turbulence models regardless of 
the presence or absence of magnetic fields. The $\Delta$-variance 
$\sigma_{\Delta}^2(n)$ exhibits a negative slope
and peaks at small scales as soon as local collapse produces dense
cores. This is in contrast to the case of non-self-gravitating
hydrodynamic turbulence where $\sigma_{\Delta}^2(n)$ has a
positive slope and the maximum at the largest scales. Our results can
therefore be used to differentiate between different
stages of protostellar collapse in star-forming molecular clouds
and to determine scaling properties of the underlying
turbulent velocity field.

The effect of protostellar collapse, however, is not visible in
molecular line maps of star-forming clouds, as all 
molecules trace only a limited dynamic range of
densities. The density contrast in star-forming regions is
much larger. ${}^{12}$CO and ${}^{13}$CO observations, for example, 
trace only the inter-core gas distribution and at best the 
outer parts of individual protostellar cores. Hence, the density
structure seen in these molecules is indistinguishable
for star-forming and non-star-forming regions.

As resolving high density contrasts is the key for
detecting the effect of star formation in the $\Delta$-variance, we
propose observations of dust continuum or of the dust extinction
instead. These techniques do not have the same limitations of the
dynamic range and are
therefore better suited to quantitatively study the full density
evolution during the star-formation process. This is confirmed by a
first comparison of our models with the 3mm dust continuum map 
taken by Testi \& Sargent (1998) in Serpens.
 
\begin{acknowledgements}
%\acknowledgments
We thank Leonardo Testi and Anneila Sargent for kindly providing us
the FITS file of their 3mm continuum map of the Serpens
core. Furthermore we are grateful to J.\ Ballesteros-Paredes, P.\
Bodenheimer, A.\ Burkert, C. Kramer, and M.-M.\ Mac Low for valuable
discussions. VO was supported by the Deut\-sche
For\-schungs\-ge\-mein\-schaft through the grant SFB 494B. RSK
acknowledges support by a Otto-Hahn-Stipendium of the
Max-Planck-Gesell\-schaft and subsidies from a NASA astrophysics
theory program supporting the joint Center for Star Formation Studies
at NASA-Ames Research Center, UC Berkeley, and UC Santa
Cruz. Computations presented here were performed at the GRAPE-Cluster
at the Max-Planck-Institut f{\"u}r Astronomie in Heidelberg, at the
Sterrewacht Leiden, at the Rechen\-zentrum Garching of the
Max-Planck-Gesell\-schaft, and at the National Center for
Supercomputing Applications (NCSA). ZEUS was used by courtesy of the
Laboratory for Computational Astrophysics at the NCSA. This research
has made use of NASA's Astrophysics Data System Abstract Service.
\end{acknowledgements}

\end{document}